\documentclass[11pt ]{article}
\usepackage{amsmath}
\usepackage{graphicx}
\usepackage{hyperref}
\usepackage{lmodern}
\usepackage{amssymb}
\usepackage{booktabs}
\usepackage{tikz}
\usetikzlibrary{patterns}
\newcommand{\lastcfrac}[2]{%
	\vphantom{\cfrac{#1}{#2}}%
	\raisebox{\dimexpr1ex-\height}{%
		$\displaystyle
		\raisebox{.5\height}{$\ddots$}+\cfrac{#1}{#2}
		$%
	}%
}
\newcommand{\lastcfracc}[2]{%
	\vphantom{\cfrac{#1}{#2}}%
	\raisebox{\dimexpr1ex-\height}{%
		$\displaystyle
		\raisebox{.5\height}{$\ddots$}
		$%
	}%
}
\newcommand{\be}{\begin{equation}}
\newcommand{\ee}{\end{equation}}
\newcommand{\beqa}{\begin{eqnarray}}
\newcommand{\eeqa}{\end{eqnarray}}

\renewcommand{\theequation}
{\arabic{section}.\arabic{equation}}

\makeatletter
\def\eqnarray{ \stepcounter{equation} \let\@currentlabel=\theequation
 \global\@eqnswtrue
 \global\@eqcnt\z@
 \tabskip\@centering
 \let\\=\@eqncr
 $$\halign to \displaywidth\bgroup\@eqnsel\hskip\@centering
 $\displaystyle\tabskip\z@{##}$&\global\@eqcnt\@ne
 \hfil$\displaystyle{{}##{}}$\hfil
 &\global\@eqcnt\tw@$\displaystyle\tabskip\z@{##}$\hfil
 \tabskip\@centering&\llap{##}\tabskip\z@\cr}
\makeatother

\makeatletter
\def\@arrayacol{\edef\@preamble{\@preamble \hskip .5\arraycolsep}}
\def\array{\let\@acol\@arrayacol \let\@classz\@arrayclassz
\let\@classiv\@arrayclassiv \let\\\@arraycr\def\@halignto{}\@tabarray}
\makeatother
\makeatletter
\newcounter{subeqncnt}
\def\thesubeqncnt{\alph{subeqncnt}}
\def\subequations{\begingroup%
   \stepcounter{equation}\edef\@tempa{\theequation}%
   \let\c@equation\c@subeqncnt\c@subeqncnt\z@
   \edef\theequation{\@tempa\noexpand\thesubeqncnt}}

\makeatother
\newcommand{\nn}{\nonumber}

\DeclareMathOperator{\arccosh}{arccosh}

\def\CC {{\cal C}}
\def\CD {{\cal D}}
\def\CE {{\cal E}}
\def\CF {{\cal F}}

\def\CM {{\cal M}}

\begin{document}

\setlength{\baselineskip}{6mm}
\begin{titlepage}
\begin{flushright}

{\tt NRCPS-HE-05-2018} \\

\end{flushright}

\begin{center}
{\Large ~\\{\it   Artin Billiard \\
 \vspace{1cm}
 
Exponential Decay of  Correlation Functions\\
 
}

}%title ends

\vspace{2.5cm}
%\author{

 {\sl  Hasmik Poghosyan \footnote{${}$ On a leave of absence  from the 
 A.I. Alikhanyan National Science
Laboratory,Yerevan,0036,Armenia}, Hrachya Babujian\footnote{${}$ On a leave of absence from the 
  A.I. Alikhanyan National Science
Laboratory,Yerevan,0036,Armenia}    and 
George Savvidy

 \bigskip
 \centerline{${}$ \sl Institute of Nuclear and Particle Physics}
\centerline{${}$ \sl Demokritos National Research Center, Ag. Paraskevi,  Athens, Greece}
\bigskip

}%author ends
%}
%\date{}%in order NOT to write the date
%\maketitle
\end{center}
\vspace{30pt}

\centerline{{\bf Abstract}}
  The hyperbolic Anosov  C-systems  have exponential instability of their  trajectories and  as such  represent the most natural  chaotic dynamical systems. Of special interest are  C-systems which are defined on compact surfaces  of  the Lobachevsky plane of constant negative curvature.  An example of such system has been introduced in a brilliant article published in 1924 by the mathematician Emil Artin.  The  dynamical system is defined on the fundamental region of the Lobachevsky plane which is obtained by the identification of points  congruent with respect to the modular group,  a  discrete subgroup of the Lobachevsky plane isometries.  The fundamental region in this case is a hyperbolic triangle.  The geodesic trajectories of  the non-Euclidean 
billiard are bounded to propagate  on the fundamental  hyperbolic  triangle.  In this article we  shall expose his results, will calculate the correlation 
functions/observables  which are defined on the phase space of the Artin billiard and demonstrate 
the exponential decay of the correlation functions with time.  We use  Artin symbolic dynamics,   the differential geometry and group theoretical  methods of Gelfand and Fomin.

\vspace{12pt}

\noindent

\end{titlepage}

\section{\it Artin Dynamical System with Quasi-Ergodic Trajectories}
In this section we shall  recall the ideas and formulas from
the brilliant article of Emil Artin \cite{Artin} published in 1924  
describing an example of ergodic dynamical system which is realised as a geodesic flow 
on a compact surface $\CF$ of the Lobachevsky plane. The aim of his article was to construct an example  of a dynamical system in which  "almost all" geodesic trajectories are quasi-ergodic 
\cite{hadamard,hedlund,anosov,bowen0,kolmo,kolmo1,sinai3,ruelle,hopf,Hopf,Gelfand,Collet,moore,dolgopyat,chernov,yangmillsmech,Savvidy:1982jk,yer1986a,konstantin,Savvidy:2015ida,Savvidy:2015jva,Maldacena:2015waa}, meaning that all trajectories, with the exception of measure zero,  during their time evolution will approach infinitely close any point and any given direction on  surface $\CF$\footnote{
 In subsequent investigations it was proven that the geodesic flow on compact surfaces of 
 constant negative curvature has mixing of all orders, Lebesgue spectrum and nonzero  Kolmogorov entropy 
 \cite{anosov,bowen0,kolmo,kolmo1,sinai3,ruelle,Hopf,Gelfand,Collet,moore}. The 
 Artin construction has an advantage of being extremely  transparent, well motivated physically and 
 appealing to the intuition.  }.
 
\begin{figure}
\centering
\begin{tikzpicture}[scale=2]
\clip (-4.3,-0.4) rectangle (4.5,2.8);
\draw[step=.5cm,style=help lines,dotted] (-3.4,-1.4) grid (3.4,2.6);
\draw[,->] (-2.3,0) -- (2.4,0); \draw[->] (0,0) -- (0,2.3);
\foreach \x in {-2,-1.5,-1,-0.5,0,0.5,1,1.5,2}
\draw (\x cm,1pt) -- (\x cm,-1pt) node[anchor=north] {$\x$};
\foreach \y in {1}
\draw (1pt,\y cm) -- (-1pt,\y cm) node[anchor=east] {$\y$};
\draw (0,0) arc (0:180:1cm);
\draw (1,0) arc (0:180:1cm);
\draw (2,0) arc (0:180:1cm);
\draw (-1,0) arc (0:95:1cm);
\draw (1,0) arc (0:-95:-1cm);
\draw (1.5,0)--(1.5,2.1);
\draw (0.5,0)--(0.5,0.86602540378);
\draw (-0.5,0)--(-0.5,0.86602540378);
\draw (-1.5,0)--(-1.5,2.1);
\draw[ ultra thick, blue] (0.5,0.86602540378)--(0.5,2.1);
\draw[ ultra thick, blue] (-0.5,0.86602540378)--(-0.5,2.1);
\draw[ultra thick, blue] (0.5,0.86602540378) arc (60:120:1cm);
\draw (-0.25,1.8) node{ $\CF$};
\draw (0,2.4) node{ $\CD$};
\draw [thin,gray](2.35,0) arc (0:180:1.5cm);
\draw[fill] (0.1,1.29903810568) circle [radius=0.02];
\draw [->,] ((0.1,1.29903810568) to ((0.2,1.36);
\draw (0.15,1.2) node[scale=0.6]{ $(x,y)$};
\draw (0.15,1.45) node[scale=0.6]{ $\vec{v}$};
\draw (0,2.4) node{ $\CD$};
\draw (-0.6,0.7) node{ $A$};
\draw (0.4,0.7) node{ $B$};
\draw (-0.1,0.85) node{ $C$};
\draw[fill] (-0.5,0.86602540378) circle [radius=0.02];
\draw[fill] (0,1) circle [radius=0.02];
\draw[fill] (0.5,0.86602540378) circle [radius=0.02];
\draw[fill] (-0.5,1.7) circle [radius=0.025];
\draw[fill] (0.5,1.7) circle [radius=0.025];
\draw (-0.6,1.7) node[scale=0.8]{ $a$};
\draw (0.6,1.7) node[scale=0.8]{ $a$};
\draw[fill] ( -0.4,0.915) circle [radius=0.025];
\draw[fill] ( 0.4,0.915) circle [radius=0.025];
\draw (-0.35,0.85) node[scale=0.8]{ $b$};
\draw (0.3,0.85) node[scale=0.8]{ $b$};
\draw [<-,thin,  teal,dashed] (1.3,1.43) to [out=90,in=120] (2,1.6) ;
\draw  [thin,  teal] (2.1,1.6) node{ $K$};
\draw[thin,teal] (-0.5,1.1)  arc (60:30:0.41cm);
\draw[thin,teal] (0.35,0.952)  arc (150:120:0.41cm);
\draw [thin,teal] (-0.43,0.98) node[scale=0.7]{ $\alpha$};
\draw [thin,teal] (0.43,0.98) node[scale=0.7]{ $\beta$};
\end{tikzpicture}
\caption{The  fundamental region $\CF$ is the hyperbolic triangle $ABD$,  the vertex D is at infinity of the $y$ axis. The edges of the triangle are the arc $AB$,  the rays $AD$   and $BD$. The points on the edges $AD$ and $BD$ and the points of the arks $AC$ with $CB$ 
should be identified by the transformations $w=z+1$ and $w = -1/z$ in order to form a {\it closed compact surface} $\bar{\CF}$
by  "gluing" the opposite edges of the modular triangle together. 
This  forms a closed compact surface $\bar{\CF}$.   The hyperbolic triangle $OAB$ equally well can be considered as the fundamental region. The modular transformations (\ref{modular}) of the fundamental  region $\CF$ create a regular tessellation of the  whole Lobachevsky plane by congruent hyperbolic triangles. 
 K is the geodesic trajectory passing through the point ($x,y$) of $\CF$ in the $\vec{v}$ direction.}
\label{fig1} 
\end{figure}
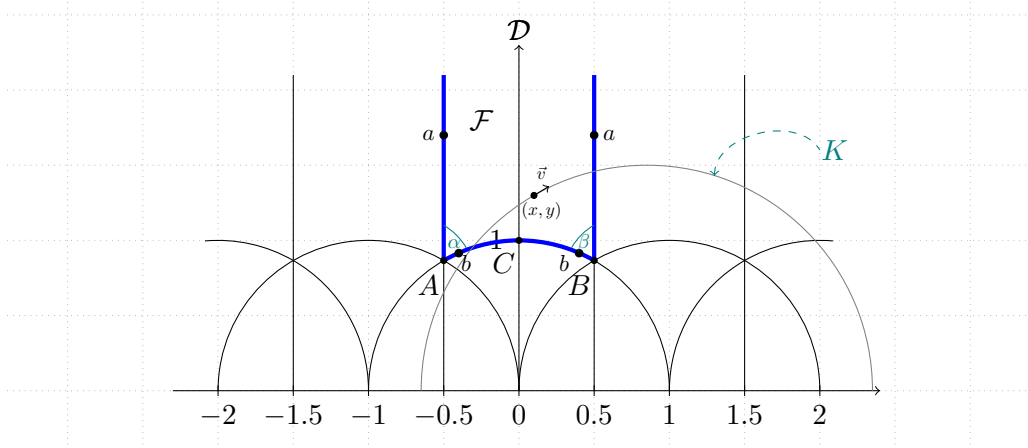

Following the Artin construction let us consider the model of the Lobachevsky plane realised 
in the upper half-plane $y>0$ of the complex plane  $z=x+iy \in \CC$
with  the Poincar\'e metric which is given by the line element \cite{Poincare}
\be
\label{metric}
ds^2 = {dx^2 +dy^2 \over y^2} = \frac{dz\,d\bar{z}}{ (\Im z)^2}\,,
\ee
where  $\bar{z}$ is the complex  conjugate of $z$ and $\Im z$ is the imaginary part of $z$.
The Lobachevsky plane is a surface of a constant negative curvature, because its  curvature is equal to $R=g^{ik}R_{ik}= -2$  and it is twice the Gaussian curvature $K=-1$.
This  metric has two well known properties: 1) it is invariant with respect to all linear substitutions, which form the group $g \in G$ of isometries of the Lobachevsky plane\footnote{$G$ is a subgroup of all M\"obius transformations. }:
 \beqa\label{real_frac_trans}
w= g \cdot z \equiv \left(
\begin{array}{cc}
\alpha  & \beta   \\
  \gamma   & \delta  
\end{array} \right) \cdot z  \equiv \frac{\alpha z +\beta}{\gamma z +\delta},  
\eeqa
where $\alpha, \beta,  \gamma, \delta $ are {\it real coefficients of the matrix} $g $ and the determinant 
of $g$ is positive, $  \alpha \delta -   \beta  \gamma  > 0 $. Indeed, 
from (\ref{real_frac_trans}) one can get
\beqa
dw=\frac{\alpha \delta-\beta \gamma}{(\gamma z+\delta)^2} dz\,, \quad
d \bar{w}=\frac{\alpha \delta-\beta \gamma}{(\gamma \bar{z}+\delta)^2} d\bar{z}\,,\quad
\Im w=y\, \frac{\alpha \delta-\beta \gamma}{(\gamma z+\delta) (\gamma \bar{z}+\delta)}\,, \nn
\eeqa 
thus
\be
\frac{dw\,d\bar{w}}{(\Im w)^2}=\frac{dz\,d\bar{z}}{(\Im z)^2}\,. \nn
\ee
The metric (\ref{metric}) is also invariant under the reflection $w=-\bar{z}$.
2) The geodesic lines are either semi-circles orthogonal to the real axis  or 
rays perpendicular to the real axis. The equation for the geodesic lines on a curved 
surface has the form 
\be
\label{eqmotion}
\frac{d^2 x^i}{d t^2}+\Gamma^i_{kl}\frac{d x^k}{d t}\frac{d x^l}{d t}=0\,,
\ee
where  $\Gamma^i_{kl}$ are the  Christoffer symbols  of  the metric  in (\ref{metric})
\be \label{metricten}
g_{ik}=
\left(
\begin{array}{cc}
\frac{1}{y^2} & 0 \\
0 & \frac{1}{y^2} \\
\end{array}
\right)~
\ee
and the geodesic equation  (\ref{eqmotion}) takes the form 
\beqa
&&\frac{d^2 x}{d t^2}-\frac{2}{ y} \frac{d x}{d t}\frac{d y}{d t}=0\,, ~~~~~
 \frac{d^2 y}{d t^2}+\frac{1}{ y}\left(\frac{d x}{d t}\right)^2-
\frac{1}{ y}\left(\frac{d y}{d t}\right)^2=0 \nn
\eeqa
and has two solutions
\beqa\label{traject}
& x(t)-x_0=r \tanh \left(t \right),~~~~  y(t)=\frac{r}{\cosh\left(t \right)}~~~&\leftarrow
\text{orthogonal semi-circles}~ \nn\\
& x(t)=x_0,~~~~~~~~~~~~~~~~~~~~~~  y(t)=e^t ~~~ &\leftarrow
\text{perpendicular rays }~,
\eeqa
the result we mentioned  above.  Here $x_0 \in (-\infty, +\infty), t \in (-\infty, +\infty)$ and $r \in (0, \infty)$.
Substituting each of the above solutions into the metric (\ref{metric}) one 
can get convinced that the points on the geodesics curves move with a unit velocity
\be\label{unit}
{ds \over dt } =1.
\ee 
In order to construct a compact surface $\CF$ on the  Lobachevsky plane, one can identify all points in the upper half of the plane which are related to each other by the substitution (\ref{real_frac_trans})  with the integer coefficients and a unit determinant. These transformations form  a modular group $d \in D$. Thus we consider two points $z$ and $w$  to be "identical"
if:
\begin{equation}\label{modular}
w =\frac{mz+n}{pz+q},~~~~
d=\left(
\begin{array}{cc}
m & n \\
p & q \\
\end{array} \right), ~~~~d \in D 
\end{equation}
with  integers $m$, $n$, $p$, $q$   constrained by the condition $mq-pn=1$.  The  $D$  is 
the discrete subgroup of the isometry transformations $G$ of (\ref{real_frac_trans})\footnote{The modular group $D$ serves as an example of the Fuchsian group \cite{Poincare,Fuchs}. Recall that Fuchsian groups are discrete subgroups of the group  of all isometry transformations $G$ of 
(\ref{real_frac_trans}). The Fuchsian group allows to tessellate the hyperbolic plane with regular polygons as faces, one of which can play the role of the fundamental region.}. 
The identification creates a regular tessellation of the  Lobachevsky plane by congruent hyperbolic triangles in Fig. \ref{fig1}.   The Lobachevsky plane is covered by the infinite-order triangular tiling. 
One of these triangles can be chosen as a fundamental region. That fundamental region $\CF$ of the above modular group (\ref{modular}) is the
well known "modular triangle", consisting of those  points between the lines
$x=-\frac{1}{2}$ and $x=+\frac{1}{2}$ which lie outside the unit circle in  Fig. \ref{fig1}.
The modular triangle $\CF$ has two 
equal angles  $\alpha = \beta = \frac{\pi }{3}$ and  with the third one equal to zero, $\gamma=0$, 
thus $\alpha + \beta + \gamma = 2 \pi /3 < \pi$.
The area  of the fundamental region is finite and equals to $\frac{\pi }{3}$. 
The invariant area element on the Lobachevsky plane is proportional to the square root of the determinant  of the matrix (\ref{metricten}):
\be\label{me}
d \Omega=\sqrt{g} dx dy = {dx dy \over y^2} \,, 
\ee
thus
\be
\text{Area}(\CF)=\int _{-\frac{1}{2}}^{\frac{1}{2}} dx
\int _{\sqrt{1-x^2}}^{\infty }\frac{dy}{y^2}  =\frac{\pi }{3}\,.\nn
\ee
Inside the modular triangle $\CF$ there is exactly one representative
among all equivalent points of the Lobachevsky plane with the exception 
of the points on the triangle edges which are opposite to each other. 
These points should be identified in order to form a {\it closed compact surface} $\bar{\CF}$
by  "gluing" the opposite edges of the modular triangle together.
On  Fig. \ref{fig1} one can see the pairs of points on the edges of the triangle  
which are identified.
This construction allows to consider the geodesic flow on the abstract  surface $\bar{\CF}$ associated with Lobachevsky plane.  
Because of the homogeneity of the Poincar\'e metric, the metric on $\bar{\CF}$  is regular, except for its vertex  points. The main goal of the  construction is to consider now the behaviour of the geodesic trajectories defined on the  surface $\bar{\CF}$ of constant negative curvature. 

In order to describe the behaviour of the geodesic trajectories on the surface $\bar{\CF}$
one can use the knowledge of the geodesic trajectories on the whole Lobachevsky plane (\ref{traject}). Let us consider an arbitrary point $(x,y) \in \CF$ and the 
velocity vector $\vec{v} = (\cos \theta, \sin \theta)$. These are the coordinates of the 
phase space $(x,y,\theta) \in \CM$, and they uniquely determine  the geodesic trajectory  as 
the orthogonal circle $K$ in the whole Lobachevsky plane.  As this trajectory "hits" the edges 
 of the fundamental region $\CF$ 
and goes outside of it,  one should apply the modular transformation (\ref{modular}) to that parts of the circle $K$ which lie  outside of $\CF$ in order to return them back to the $\CF$. 
That algorithm will define the whole trajectory on $\bar{\CF}$ for $t \in (-\infty, +\infty)$.

One should observe that this description of the trajectory on $\bar{\CF}$ is equivalent to the 
 set of geodesic circles $\{ K' \}$ which appear under the action of the 
modular group (\ref{modular}) on the initial  circle $K$.  One should join together 
the parts of the geodesic circles $\{ K' \}$  which 
lie  inside $\CF$ into a unique continuous trajectory on $\bar{\CF}$ (see Fig.\ref{fig4}).   
In this context  the quasi-ergodicity of the trajectory $K$ on $\bar{\CF}$ will mean that among  all geodesic circles 
$\{ K' \}$  there are those which  are approaching arbitrarily close to any given circle $C$.
Now notice that the geodesic circle $K$ is determined by its  {\it base points} $\xi$ and $\eta$,
which lie on the real axis.  Under the action of the modular group (\ref{modular})  
the coordinates  $\xi$  and $\eta$  will be mapped  into the 
base points $\xi'$ and $\eta'$ of the transformed circle $K^{'}$:
\be\label{modulapointtran}
\xi' =\frac{m\xi+n}{p\xi+q},\qquad
\eta' =\frac{m\eta+n}{p\eta+q}~.
\ee
In this context the geodesic circles can  be considered  "close" to each other if their base points lie in the infinitesimal neighbourhood. 
  
It is convenient to introduce  the plane $\CE$ with the coordinates $(\xi, \eta) \in \CE$.
To each point ($\xi$, $\eta$) of the $\CE$  plane corresponds a geodesic circle on the z-plane, 
with $\xi$, $\eta$ being its base coordinates. And conversely to every circle of the $z$-plane 
one  can assign  two  points ($\xi$, $\eta$) or ($\eta$, $\xi$)  on the $\CE$ plane. 
The reason for this ambiguity lies in the fact that the ordering of the base point coordinates 
after the action of  the modular transformation can be inverted.  The geodesic circles 
were considered to be in the infinitesimal neighbourhood 
 if their base coordinates were close.  
 This implies now that the points ($\xi' = d \xi$, $\eta' =d  \eta$)  on the $\CE$ plane resulting 
 from  the  action of the full group of modular transformation (\ref{modulapointtran}) on ($\xi$, $\eta$) 
 must be everywhere dense on the $\CE$ plane if the  
trajectory $K$ is quasi-ergodic. 
\begin{figure}
\centering
\begin{tikzpicture} [scale=0.60]
\draw[step=1cm,gray,very thin,dotted] (-3.9,-3.9) grid (7.9,5.9);
\draw[very thick,->] (-2.5,0) -- (6.5,0) node[anchor=north west] {$\xi$ axis};
\draw[very thick,->] (0,-3.5) -- (0,4.5) node[anchor=south east] {$\eta$ axis};
\foreach \x in {-2,-1,.,1,2,3,4,5,6}
\draw (\x cm,1pt) -- (\x cm,-1pt) node[anchor=north] {$\x$};
\foreach \y in {1,2,3,4}
\draw (1pt,\y cm) -- (-1pt,\y cm) node[anchor=east] {$\y$};
\path[pattern=north east  lines, pattern color=blue] (-2,-3.3) rectangle (6.2,-3);
\path[pattern=north west  lines, pattern color=blue] (-1,-3) rectangle (6.2,-2);
\path[pattern=north east  lines, pattern color=blue] (0,-2) rectangle (6.2,-1);
\path[pattern=crosshatch dots, pattern color=blue] (1,-1) rectangle (6.2,0);
\path[pattern=north east  lines, pattern color=blue] (2,0) rectangle (6.2,1);
\path[pattern=north west  lines, pattern color=blue] (3,1) rectangle (6.2,2);
\path[pattern=north east  lines, pattern color=blue] (4,2) rectangle (6.2,3);
\path[pattern=north west  lines, pattern color=blue] (5,3) rectangle (6.2,4);
\path[pattern=north east  lines, pattern color=blue] (6,4) rectangle (6.2,4.3);
\draw[dashed] (-2.2,-2.2) --(4.3,4.3);
\draw (-1.5,3.5) node{ $\CE$};
\draw (4.5,4.5) node{ $\xi=\eta$};
%\draw[green,thick, ->] (-2.7,1.8) .. controls (2,3) and (4,0) .. (1.5,-0.5);
%\draw [black,thick] (-2.7,1.5) node{ $-1<\eta<0, \xi>1$};
%\draw [green, very thick] (-4.9,1.2) rectangle (-0.5,1.8);
\end{tikzpicture}
\caption{The  $\CE$ plane.  For a trajectory to be a quasi-ergodic 
it is necessary and sufficient to have an everywhere dense distribution of  the 
points ($\xi'$, $\eta'$) (\ref{modulapointtran})
in the subregion of  the $\CE$ plane defined by the conditions 
$-1<\eta<0\, ,  1<\xi $. The region between the "stairs" and the diagonal $\xi = \eta$
corresponds to the geodesic circles which are not crossing the fundamental region $\CF$.}
\label{fig2} 
\end{figure}
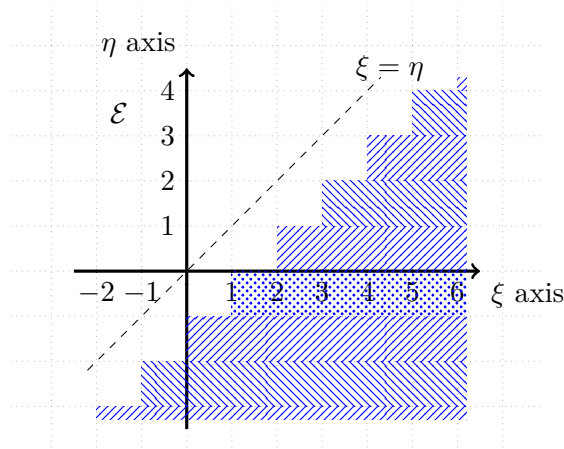

In the Artin  article \cite{Artin} it was shown  that in order to have quasi-ergodic behaviour of a trajectory 
it is necessary and sufficient to have an everywhere dense distribution of  points ($\xi'$, $\eta'$)
in the subregion of  the $\CE$ plane defined in  Fig. \ref{fig2} by the condition
\be \label{region}
-1<\eta<0\, , \qquad 1<\xi\, .
\ee
The necessity of this condition is obvious. Provided this condition is fulfilled,
one can apply to the points of this region the modular transformations
\be\label{region1}
\xi'=\xi+n\,, \qquad
\eta'=\eta+n\,
\ee
with some integer $n$ and to see that the areas
$n<\xi$, $n-1<\eta<n$ are covered everywhere dense. These are the regions under the 
"stairs"  shown in  Fig. \ref{fig2}. 
One can fix the arbitrariness of the base points coordinates, mentioned above, by   
ordering them as  $\eta<\xi$.   It follows  then that it is sufficient  to require 
an everywhere dense distribution 
of points  on the $\CE$ plane for the geodesic circles which are crossing   the fundamental 
region $\CF$.  That set of points of the $\CE$ plane lie exactly inside the region defined above by equations (\ref{region}) and (\ref{region1}).
Thus one can assume that the base point coordinates of the initial  circle $K$  
lie in the region $-1<\eta<0$,~ $1 < \xi $.

The coordinates $\xi$ and $-\eta$ in that region can be represent  as the continued fractions  with   positive integers $a_i$
\beqa \label{contfrac}
\xi = 
a_0+\cfrac{1}{a_1 +\cfrac{1}{a_2 +\cfrac{1}{a_3 + \lastcfracc{arg1}{arg2} }}}\,,
\qquad
-\eta = \cfrac{1}{a_{-1} +\cfrac{1}{a_{-2} +\cfrac{1}{a_{-3} + \lastcfracc{arg1}{arg2} }}}\,,
\eeqa
{\it which  means that the geodesic trajectory $K$ on $\bar{\CF}$ can be represented as an infinite} $A$-chain
of the form 
\be\label{secuence}
......., a_{-3}, a_{-2}, a_{-1}, a_0, a_{1},a_{2},a_{3}, ......
\ee
Let us define  for any  integer $n\geq 0$ the coordinates 
$ \eta_n$ and $\xi_n $ by the relations 
\beqa \label{desfrac}
\xi_n = a_n+\cfrac{1}{a_{n+1} +\cfrac{1}{a_{n+2} +\cfrac{1}{a_{n+3} + \lastcfracc{arg1}{arg2} }}}\,,
\qquad
-\eta_n = \cfrac{1}{a_{n-1} +\cfrac{1}{a_{n-2} +\cfrac{1}{a_{n-3} + \lastcfracc{arg1}{arg2} }}}\,\,.
\eeqa
This corresponds to the shift on the $A$-chain expansion to the right. 
These coordinates satisfy the relations
\beqa \label{recrel}
\xi =\frac{P_n\xi_n +P_{n-1}}{Q_n \xi_n +Q_{n-1}}\,,
\qquad
\eta =\frac{P_n\eta_n +P_{n-1}}{Q_n \eta_n +Q_{n-1}}\,,
\eeqa
where $P_n$ and $Q_n$ are positive integers. Indeed, 
from (\ref{contfrac}) and (\ref{desfrac}) we obtain  
\be
\xi=a_0 +\frac{1}{\xi_1}=\frac{a_0\xi_1+1}{\xi_1}\,,
\qquad
\eta=a_0+\frac{1}{\eta_1}=\frac{a_0\eta_1+1}{\eta_1}\,,
\ee
so that from (\ref{recrel}) we see that
\beqa \label{P_0,Q_0}
&&P_1=a_0\,, \qquad P_0=1\,, \\
&&Q_1=1\,, \qquad Q_0=0\, . \nonumber
\eeqa
Now one can use the mathematical induction.  Suppose  that  (\ref{recrel}) is true for some $n$, then (\ref{desfrac}) can be rewritten as
\be \label{recrel2}
\xi_n=a_n+\frac{1}{\xi_{n+1}}\,,
\qquad
\eta_n=a_n+\frac{1}{\eta_{n+1}}\,,
\ee
and then inserting this equations into the (\ref{recrel})  we shall get
\beqa
\xi=
\frac{P_n\left(a_n+\frac{1}{\xi_{n+1}}\right)+P_{n-1}}
{Q_n\left(a_n+\frac{1}{\xi_{n+1}}\right)+Q_{n-1}}
=
\frac{\left(P_na_n+P_{n-1}\right)\xi_{n+1}+P_n}
{\left(Q_na_n+Q_{n-1}\right)\xi_{n+1}+Q_n}\,,\\
\eta=
\frac{P_n\left(a_n+\frac{1}{\eta_{n+1}}\right)+P_{n-1}}
{Q_n\left(a_n+\frac{1}{\eta_{n+1}}\right)+Q_{n-1}}
=
\frac{\left(P_na_n+P_{n-1}\right)\eta_{n+1}+P_n}
{\left(Q_na_n+Q_{n-1}\right)\eta_{n+1}+Q_n}\,.
\eeqa
By defining 
\beqa
\label{PQrel}
&&P_{n+1}=P_na_n+P_{n-1}\,,\\ \nonumber
&&Q_{n+1}=Q_na_n+Q_{n-1}\,
\eeqa
we shall complete the prove of the equations (\ref{recrel}). The sequences $P_n$ and $Q_n$ satisfy  the relation
\begin{equation} \label{PQrecrel}
P_n Q_{n-1} - Q_n P_{n-1} =(-1)^n~.
\end{equation}
Indeed, using   (\ref{PQrel})
we shall get the recurrence relation:
\be \label{PQRECREL}
P_nQ_{n-1}-Q_nP_{n-1}=-\left(P_{n+1}Q_n-Q_{n+1}P_n\right)\,.
\ee
It follows from (\ref{P_0,Q_0}) that for  $n=1$ we have
\be \label{PQRECREL00}
P_1Q_0-Q_1P_0=-1.
\ee
Now  (\ref{PQrecrel}) is an obvious consequence of (\ref{PQRECREL}) and (\ref{PQRECREL00}). 
Let us consider the integer matrices  constructed in terms of 
 $P_n$ and $ Q_n$ as 
\be\label{inver}
d_n= \left(
\begin{array}{cc}
P_n & P_{n-1} \\
Q_n & Q_{n-1} \\
\end{array} \right). 
\ee
Because of the (\ref{PQrecrel}) the determinant of $d_n$ is equal to $(-1)^n$ and for even $n$ 
they represent the matrices  of the modular transformations (\ref{modular}). It follows 
therefore from (\ref{recrel}) and (\ref{inver}) that the points  $(\xi_n,\eta_n)$ appear under the action of the modular 
transformations $d^{-1}_n$ on the point $(\xi,\eta)$ 
\be\label{recurr}
\xi_n = d^{-1}_n \cdot \xi,~~~\eta_n = d^{-1}_n \cdot \eta~.
\ee 
The geodesic circle $K$ was given by its base coordinated $(\xi,\eta)$ and, as 
it was just demonstrated, the points   $(\xi_n,\eta_n)$
 are the base coordinates of the geodesic circles $\{K '\}$. 

In order for the base points coordinates of the circles $\{K '\}$ to be everywhere dense in $\CE$ it is  necessary that the continued fraction matrices  $d_n$ (\ref{inver})
for some  $n$ will match  with an arbitrary accuracy any given element of 
the modular group $D$ in (\ref{modular}). 

Let us formulate this statement in terms of sequences. 
Suppose that  a  sequence  of $2m+1$ positive numbers  
\be\label{any}
c_{-m},\dots, c_{-2}, c_{-1}, c_0,c_1,\dots, c_m
\ee 
approximate  a given circle $C$  with a sufficient accuracy,
then the circles  $\{ K' \}$ will closely approach the circle $C$ if it will 
 be possible to find an even index $n$  such that {\it the section of the $A$-chain }
(\ref{secuence})  of the length $2m+1$  
$$
a_{n-m},a_{n-m+1},\dots , a_{n-1},a_{n},a_{n+1}\dots, a_{n+m}
$$ 
will coincide  either with 
$c_{-m},\dots, c_{-2}, c_{-1}, c_0,c_1,\dots, c_m$ or with its
reverse sequence. Therefore for the quasi-ergodicity of the trajectory $K$, 
 represented  by  the infinite $A$-chain (\ref{secuence}),  it is necessary and 
sufficient to have 
all imaginable finite sequences  (\ref{any}) of positive integers to be a section 
of the  $A$-chain (\ref{secuence}).

Let us briefly outline what has been achieved.  The  geodesic trajectory $K$ is represented 
by an infinite $A$-chain (\ref{secuence}). The points on the $\CE$ plane which correspond to the circles $\{K'\}$ are generated by the algorithm  (\ref{desfrac}) and (\ref{recurr}).
In order for the geodesic trajectory $K$ to be quasi-ergodic it is necessary and sufficient
that every imaginable finite sequence of positive integers can be found as a section in 
the associated A-chain  (\ref{secuence}). In accordance with the results of  
Burstin \cite{Burstin,Minkowski} almost all numbers $\xi$ have a quasi-ergodic continued fractions.
Thus almost all  geodesic trajectories on the surface $\bar{\CF}$  are quasi-ergodic.

The fact that such chains really exist  is easy to see. Let us consider 
the lattice points of the $N$-dimensional Euclidean space with positive integer 
coordinates and  the diagonal planes passing through the vertices of the lattice. 
If one considers the coordinates of a lattice vertex written in a fixed order,
one can  obtain a sequence  $a_1,...,a_N$  of positive numbers with desired property,  
because they naturally represent a possible section of the $A$-chain of the length $N$. 
All possible chains of the length $N$ will be generated by considering the lattice points on the 
diagonal planes and steadily increasing the sum $a_1+...+a_N$  of their 
coordinates from zero to infinity.  
Increasing further $N$ one can get all possible chains of any length $a_0,a_1,a_2,\dots$. 
The numbers $a_{-1},a_{-2},\dots$ can also be generated in a similar way.
It follows that this construction of the "quasi-ergodic chains" 
has  cardinality  of the continuum.

In \cite{Artin} Artin provides examples of trajectories which are not quasi-ergodic 
 but exist in the considered system and have interesting properties. Let us consider these examples. The periodic $A$-chains correspond to the periodic orbits and vice versa.
An example of a periodic $A$-chain with a period  $3$ is given below:
\be
\dots,\, \,a_0,\,a_1,\,a_2,\,a_0,\,a_1,\,a_2,\,a_0,\,a_1,\,a_2,\,\dots
\ee
If a $A$-chain eventually becomes periodic to the right and also periodic to the left, 
then it is a "double" asymptotic trajectory.
The trajectory  "detaches" itself from a periodic trajectory to eventually approach a second (equal or different) periodic trajectory.
An example of a  $A$-chain which becomes periodic   in left with a period $3$  and to the right with period $2$  is given below:
\be
\dots,\, \,a_0,\,a_1,\,a_2,\,a_0,\,a_1,\,a_2, g,e,f,\dots,d,h,\,b_0,\,b_1,\,b_0,\,b_1,\,\dots
\ee
 Let us now consider an example of a trajectory which is not quasi-ergodic but is "quasi-periodic",
that is,  every single sector in the $A$-chain occurs, without periodicity, infinitely often.
Such a chain can be constructed, for example, starting from a quasi-ergodic 
$A$-chain and  then duplicating sectors of the $A$-chain.
The corresponding orbit then approximates itself without being quasi-ergodic, because  
 every single sector is repeated infinitely, often with an arbitrarily accuracy. 
An explicit example can be  
\be
..a_{-2}, a_{-1},a_0,a_1,a_0, a_{-1},a_0,a_1,a_2,a_{-1},a_0,a_1,{\bf a_0 }, a_{-1},a_0,a_1,a_{-2}, a_{-1},a_0,a_1,a_0, a_{-1},a_0,a_1,a_2,..\nn
\ee
Let us consider the orbit constructed in the previous example 
and immerse  a number  one in some  place of that $A$-chain.  That part of the 
chain will be never repeated, and the orbit will approach closer  and closer 
to the type of the orbit constructed above. 	
 \begin{figure}
\begin{center}
\includegraphics[width=6cm]{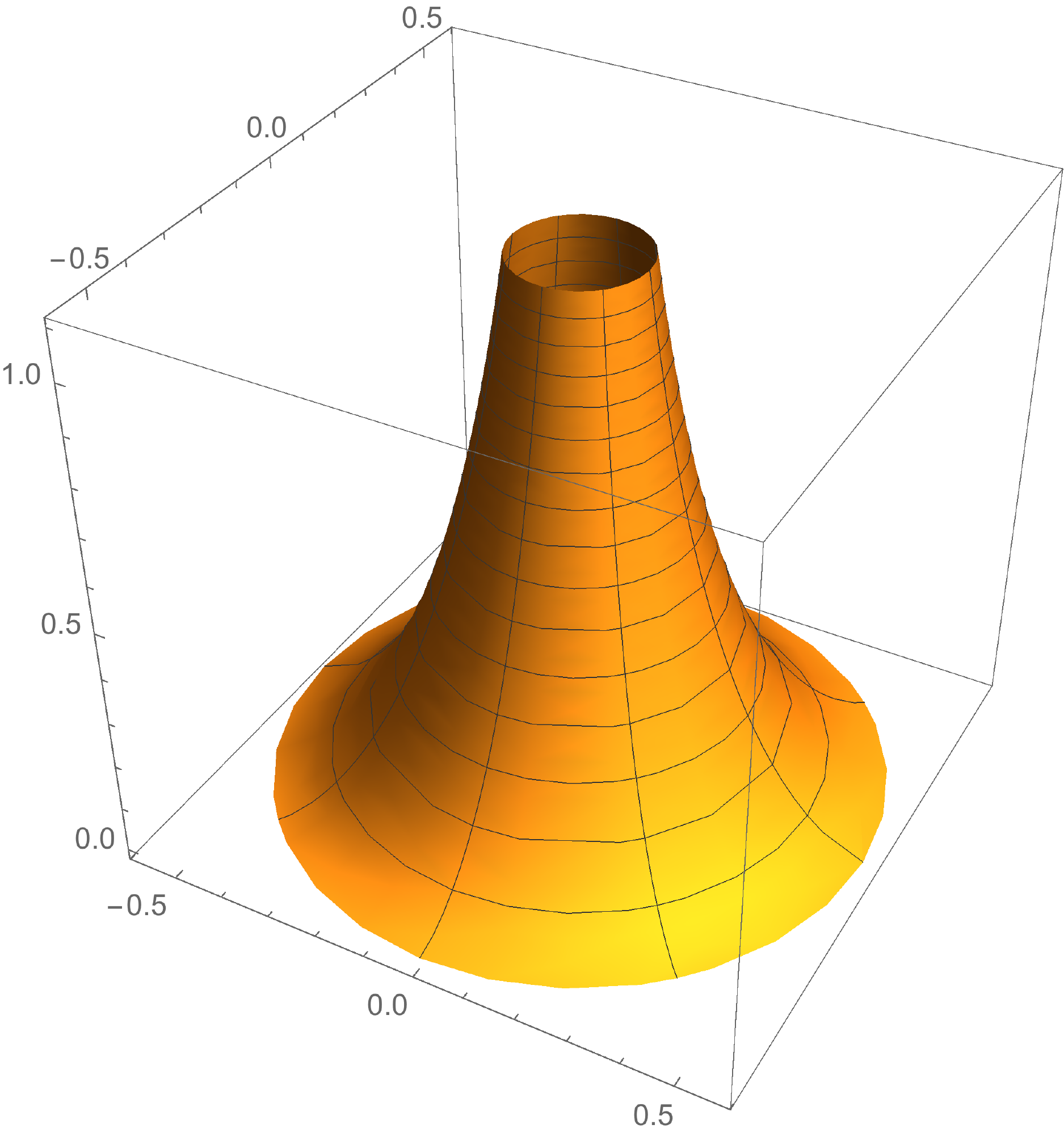}~~~
\caption{ The region $-1/2 \leq x < 1/2,~~1/2 < y$  on the Lobachevsky plane 
can be isometrically mapped  into the
surface of the pseudosphere  by the formula $u= \cos(4 \pi  x)/ \sinh(\arccosh(2 y)), ~
v = \sin(4 \pi  x)/ \sinh(\arccosh(2 y)),~  z= \arccosh(2 y) - \tanh((\arccosh(2 y)) $. The 
figure shows the image of the pseudosphere which corresponding to the $-1/2 \leq x < 1/2,~~1/2 < y$ region. 
}
\label{fig3}  
\end{center}
\end{figure}

There is more to be said about the physical realisation of the above system \cite{Artin}.
One can obtain an alternative system, if in addition to the points identified by the modular transformations, identify the points which are transforming into each another by the 
parity transformation  $z_1=-\bar{z}_2$.
The fundamental region $\CF$ is then divided  into two pieces, one of which 
is located between $x=0$ and $x=\frac{1}{2}$.  The geodesic trajectories are simply the circles. 
Since  even more points are considered to be identical the previous considerations remain valid, so that here too almost all geodesic trajectories are quasi-ergodic.

Now let us consider the pseudosphere, the surface of so called tractricoid,  shown   on  Fig. \ref{fig3}, the result of revolving a tractrix about its asymptote.  As it is known, its curvature is $K=-1$ and therefore is locally isometric to the 
 Lobachevsky plane.   One can map isometrically 
 the  fundamental region $\CF$ into the pseudosphere, so that it can be "fully settled"  on the pseudosphere.  This provides a physical realisation of 
the Artin abstract dynamical system in the ordinary three-dimensional space, because the geodesic trajectories can  be interpreted now as the movement of a mass point on the nonsingular 
part of the pseudosphere. 

\section{\it  Construction  of Periodic Geodesic Trajectories }

\begin{figure}
\begin{center}
\includegraphics[width=12cm]{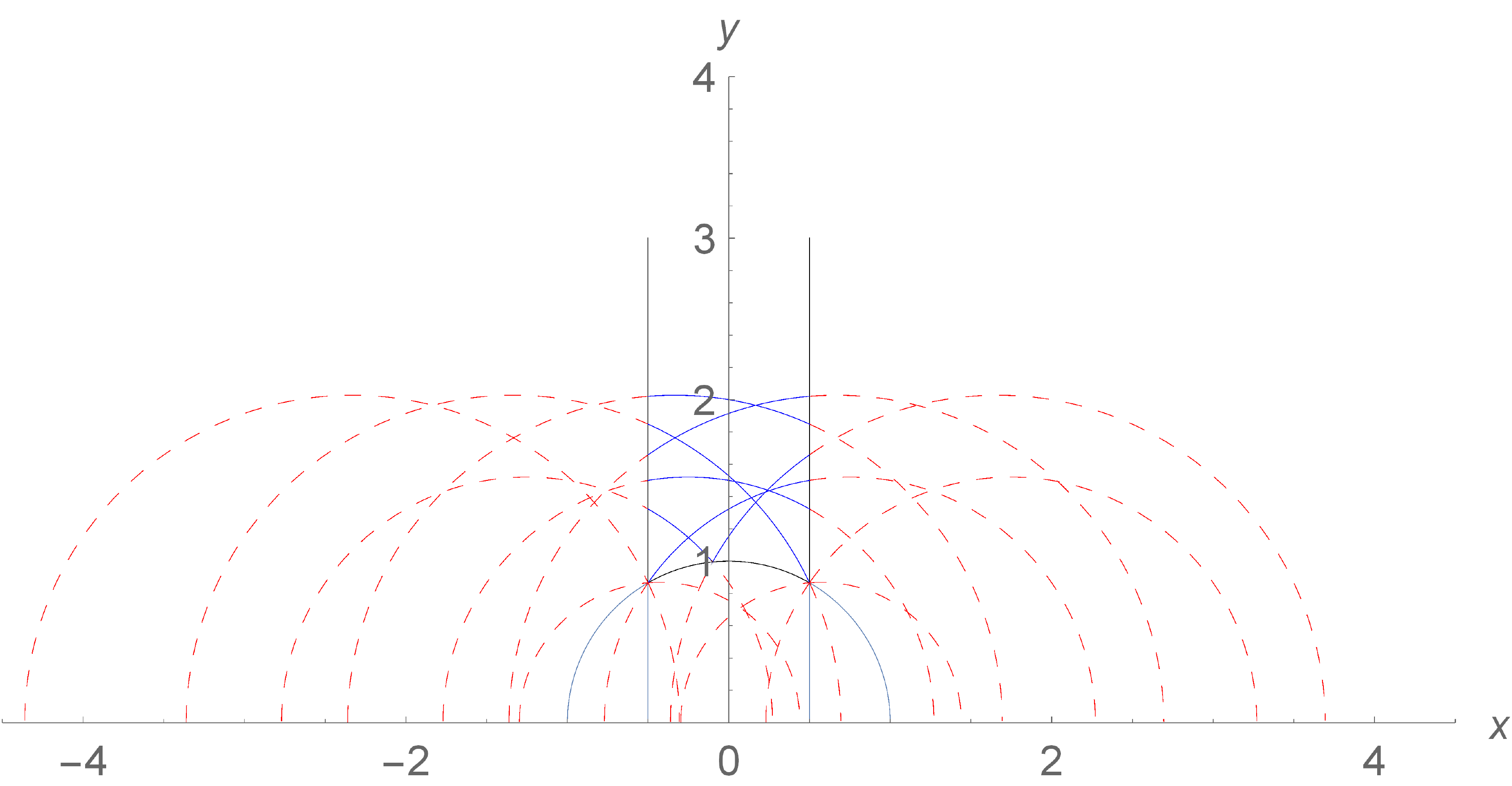}~~~
\caption{  The coordinates $(\xi,\eta)$  uniquely determine  the geodesic trajectory  as 
the orthogonal circle $K$ in the whole Lobachevsky plane.  As the trajectory "hits" the edges 
 of the fundamental region $\CF$ 
and goes outside of it,  one should apply the modular transformation (\ref{modular}) to that parts of the circle $K$ which lie  outside of $\CF$ in order to return them back to the $\CF$.  These are the 
geodesics  $\{K' \}$ which define the whole trajectory now on $\bar{\CF}$ for all the time
 $t \in (-\infty, +\infty)$.
The periodic trajectory $K$ in $\CF$ is indicated by the blue colour. The set of geodesics  $\{K' \}$ is 
shown by the red colour circles.  
}
\label{fig4}  
\end{center}
\end{figure}

In this section we shall demonstrate  how one can use  the Artin algorithm to construct a periodic 
geodesic trajectory.  Let us consider an example of trajectory which corresponds to  the periodic  $A$-chain of the form
\be
......., 1\,,2\,,3\,,1\,,2\,,3\,,1\,,2\,,3\,,1\,,2\,,3\,,1\,,2\,,3\,,1\,,2\,,3\,, ......,
\ee
where $a_0=1$.
With the help of (\ref{contfrac}) %(\ref{desfrac}) 
we can find the base points of the trajectory corresponding to the chin given above in the following form:
\beqa
\xi =1 + \cfrac{1}{2 +\cfrac{1}{3+\cfrac{1}{\xi} }}\,, \qquad
-\eta = \cfrac{1}{3 +\cfrac{1}{2+\cfrac{1}{1-\eta} }}\,,
\eeqa
where we used the fact that the chain has period three and that the continued fractions 
repeat themselves after three steps.  These are the quadratic equations on the base coordinates  
$\xi, \eta$ of $K$:
\beqa 
7 \xi^2-8\xi-3=0\,,\nn\\
7 \eta^2-8 \eta-3=0\,. \nn
\eeqa
In order to have the base points of the circle $K$ in the region (\ref{region}) we have to choose the solutions:
\be
\xi=\frac{1}{7}\left(4+\sqrt{37}\right)\,,\qquad 
\eta=\frac{1}{7}\left(4-\sqrt{37}\right)\,. 
\ee
From (\ref{real_frac_trans}),  (\ref{inver})  and (\ref{recurr}) it follows that $\xi= d_n\, \cdot\,  \xi_n$  and 
$\eta= d_n\, \cdot\, \eta_n $ therefore 
\beqa
 \xi_n= d_n^{-1} \cdot \,\xi\,,~~~~~~~ \eta_n= d_n^{-1} \cdot \,\eta\,, \nn
\eeqa
where
\be
d_n^{-1} = \begin{pmatrix}
Q_{n-1} & - P_{n-1} \\
-Q_n & P_n
\end{pmatrix}\,.\nn
\ee
These are the matrices of the modular group which are defining  the geodesic  circles
$\{K'\} $.  With the help of the last two expressions it is easy to find their  base points 
describing the periodic geodesic:   
\beqa
&\xi_0=\xi=\frac{1}{7}\left(4+\sqrt{37}\right)\,, \quad
\xi_1=\frac{1}{4} \left(\sqrt{37}+3\right)\,, \nn \\
&\xi_2=\frac{1}{3} \left(\sqrt{37}+5\right)\,, \quad
\xi_3=\frac{1}{7} \left(\sqrt{37}+4\right)\,,\quad
\xi_4= \frac{1}{3} \left(\sqrt{37}+5\right)\,
\eeqa
and
\beqa
&\eta_0=\eta=\frac{1}{7}\left(4-\sqrt{37}\right)\,, \quad
\eta_1=\frac{1}{4} \left(3-\sqrt{37}\right)\,, \\\nonumber
& \eta_2= \frac{1}{3} \left(5-\sqrt{37}\right)\,, \quad
\eta_3=\frac{1}{7} \left(4-\sqrt{37}\right)\,, \quad
\eta_4=\frac{1}{4} \left(3-\sqrt{37}\right)\,. 
\eeqa
These base coordinates define the full trajectory on the surface $\bar{\CF}$  which is depicted 
on  Fig.\ref{fig4}

\section{\it Geodesic Flow and Physical Observables on  $\bar{\CF}$  }

In the previous sections the geodesic flow on the surface $\bar{\CF}$ was formulated in 
terms of group-theoretical transformations. Here we shall use a similar approach 
in order to describe  the time 
evolution of the functions which are defined on the phase space  $(x,y,\theta) \in \CM$, where 
$z =x+iy \in \bar{\CF}$ and $\theta \in S^1$ is a
direction of a unit velocity vector.   In other words, we are interested in describing the 
time evolution of 
the physical observables $\{f(x,y, \theta)\}$.

Let us first describe a continuous time evolution of geodesics on the phase space $\CM$. 
The simplest motion on  the ray  $CD$ in Fig.\ref{fig1},  
  is given by  the solution (\ref{traject})
$x(t)=0,~~  y(t)=e^{t} $ and can be represented as a group transformation (\ref{real_frac_trans}):
\beqa\label{transformation1}
  z_1(t) =g_{1}(t) \cdot i  = \left(
\begin{array}{cc}
e^{t/2} & 0 \\
0 & e^{-t/2} \\
\end{array} \right) \cdot i  = i e^{t}~,~~~  t \in (-\infty, +\infty).
\eeqa
The other motion on the circle of a unit radius, the arc $ACB$ on  Fig.\ref{fig1}, is given by the transformation 
\beqa\label{transformation2}
  z_2(t) =g_{2}(t) \cdot i  = \left(
\begin{array}{cc}
\cosh(t/2) & \sinh(t/2) \\
\sinh(t/2) & \cosh(t/2) \\
\end{array} \right) \cdot i  = {i \cosh(t/2) +  \sinh(t/2) \over i \sinh(t/2) + \cosh(t/2)} .
\eeqa
Because the isometry group $G$ acts transitively on the Lobachevsky plane, any  geodesic 
can be  mapped into any other  geodesic through the action of the  group element $g \in G $ (\ref{real_frac_trans}), thus 
the  generic trajectory can be represented  in the following form: 
\be\label{transformation3}
z(t)= g g_{1}(t) \cdot i = \begin{pmatrix}
  \alpha e^{t/2}  & \beta e^{-t/2} \\
  \gamma e^{t/2} & \delta e^{-t/2}
 \end{pmatrix} \cdot  i , ~~~z(t)= { i  \alpha e^{ t} + \beta  \over i \gamma e^{ t} +  \delta }~.
\ee
This provides a convenient description of the time evolution of the geodesic flow on the whole Lobachevsky plane with a unit velocity vector (\ref{unit}). 

In order to project the motion into  the
 fundamental region $\bar{\CF}$ one should identify the group elements $g \in G$ which are 
 connected by the modular transformations $D$. For that one should have 
a more  convenient parametrisation of the group elements  $g \in G$ \cite{Gelfand}:
 \be\label{elements}
g = \begin{pmatrix}
  g_{11}  & g_{12}   \\
 g_{21}   &  g_{22} 
 \end{pmatrix}, ~~~~~ det g = g_{11}   g_{22} -  g_{12} g_{21}     =1.
\ee
This can be achieved by introducing a pair of complex parameters 
\be\label{para}
\omega_1 = g_{11}  +i  g_{12} ,~~~~\omega_2 = -g_{21}  -i   g_{22}
\ee 
with the determinant  condition in the following form:
\begin{equation}\label{determinant}
\Im(\overline{\omega_1} \omega_2 ) =-1.
\end{equation}
The geodesic motions induced by two elements $g$ and $g'$ of the group $ G$ 
should be identified if they are connected by the transformation $d$ of the 
modular  group $D$:
\be
g^{'}= d~ g = \left(
\begin{array}{cc}
m & -n \\
p & -q \\
\end{array} \right) \begin{pmatrix}
  g_{11}  & g_{12}   \\
 g_{21}   &  g_{22} 
 \end{pmatrix}= \left(
\begin{array}{cc}
mg_{11} - n  g_{21} & m g_{12} - n  g_{22}\\
pg_{11} - q  g_{21} & p g_{12} - q  g_{22} \\
\end{array} \right), \nn
\ee
therefore the identification takes the form  
\be
\omega'_1 = m \omega_1 + n \omega_2 ,~~~~\omega'_2 = p \omega_1 + q \omega_2,
\ee
and if one introduces  the ratios 
\begin{equation}\label{ratios}
\tau =\frac{\omega_1}{\omega_2},~~~~\tau' =\frac{\omega'_1}{\omega'_2}
\end{equation}
then one can observe that 
\be\label{modtrans}
\tau' = \frac{\omega'_1}{\omega'_2} = \frac{m \omega_1 + n \omega_2}{p \omega_1 + q \omega_2} = \frac{m \tau + n  }{p \tau + q  }, ~~~\omega'_2 =  (p \tau + q)\omega_2 .
\ee
The transformation of $\tau$  under the action of the modular group $D$ is in the 
fundamental representation. This  important fact means
that  to describe the motion on $\bar{\CF}$ the parameter $\tau = x+i y $  should belong to the fundamental region $\CF$ of the 
modular group $D$. Thus instead of describing 
the group element $g \in G$ by the pair $(\omega_2, \omega_1)$ it is more convenient to use 
 the pair $(\omega_2, \tau)$. These  parameters are connected by the determinant condition (\ref{determinant})
\be
(\tau - \bar{\tau} )\vert \omega_2 \vert^2 =2i
\ee
and any element $g$ can be defined by the parameters   $(\omega_2, \tau)$
\be\label{phasespace}
(\omega_2, \tau), ~~~\tau \in \CF ,~~~~\omega_2 = \sqrt{{2 i \over \tau - \bar{\tau}}} e^{i \theta},~~~
0 \leq \theta \leq 2 \pi,
\ee
where $\tau=x+iy$ are the coordinates in the fundamental region $\CF$ and the 
angle $\theta $ defines the 
direction of the unit velocity vector $\vec{v} = (\cos \theta, \sin \theta)$  at the point $(x,y)$ (see Fig.\ref{fig1}).  
What was achieved is that the functions on the phase space can be written 
as depending on  ($\omega_2, \tau$)  and the invariance of  functions  
with respect to the modular transformations  (\ref{modtrans}) takes the form\cite{Poincare,Poincare1}\footnote{This defines the automorphic functions, the generalisation of the trigonometric, hyperbolic, elliptic  and other periodic functions \cite{Ford}.}
\be\label{periodic}
f(\omega'_2, \tau') = f[ (p \tau +q) \omega_2, { m \tau+ n \over p\tau +q } ] = f (\omega_2, \tau).
\ee
In order to describe the evolution of the functions/observables  under the geodesic flow 
(\ref{transformation1})-(\ref{transformation3}) we have to know the parameterisation of the 
group element $g g_t$.  The pair of group elements one can  parametrised as  
 $ g \rightarrow (\omega_1,\omega_2) $  and  $g_t \rightarrow (\chi_1 ,\chi_2) $: 
\begin{eqnarray}
&\omega_1 = g_{11} +ig_{12} ,~~~~ &\chi_1= h_{11} +i h_{12} \nn\\
& \omega_2 = -g_{21} -i g_{22}  ,~~~~ &\chi_2 =-h_{21} - ih_{22}.\nn
\end{eqnarray}
The group multiplication  
defines  $ g g_t \rightarrow (\omega'_1 ,\omega'_2) $:
\begin{eqnarray}
 \omega'_1 = g_{11}h_{11} - g_{12} h_{21} + i(g_{11} h_{12} -g_{12} h_{22} )  \nn\\
 \omega'_2 = -g_{21} h_{11} + g_{22} h_{21} +i( -g_{21} h_{12} + g_{22}h_{22}), \nn
\end{eqnarray}
and the result should be expressed in terms of $(\omega_1,\omega_2) $  and  
$ (\chi_1 ,\chi_2)$ parameters:
\begin{eqnarray}
 \omega'_1 =\frac{ \omega_1 + \overline{\omega_1}}{2}  \chi_1  +
 \frac{\omega_1-\overline{\omega_1}}{2i} \chi_2              \nn \\
  \omega'_2 =\frac{\omega_2 + \overline{\omega_2}}{2} \chi_1  +
  \frac{\omega_2-\overline{\omega_2}}{2i}\chi_2 .
\end{eqnarray}
Using the above multiplication law one can find the transformation which is 
induced by the geodesic flow $g_1(t)$,  for which $\chi_1 = e^{t/2}$,   $\chi_2 = i e^{-t/2}$,
thus
\be
\omega'_1 =\cosh( t/2 ) \omega_1 + \sinh( t/2 ) \overline{\omega_1},~~~ ~
\omega'_2 =\cosh( t/2 ) \omega_2  + \sinh( t/2 ) \overline{\omega_2}
\ee 
and therefore  we shall get 
 \be\label{evolu1}
 \tau' = {\omega'_1 \over \omega'_2} = {\cosh( t/2 ) \omega_1 + \sinh( t/2 ) \overline{\omega_1} \over \cosh( t/2 ) \omega_2  + \sinh( t/2 ) \overline{\omega_2}}=
 {\tau \cosh( t/2 )  + u ~\overline{\tau} ~\sinh( t/2 )  \over \cosh( t/2 )   + u ~\sinh( t/2 ) }.
 \ee 
 Because $\tau' = x'+i y'$ and   $ u={\overline{\omega_2} / \omega_2}  
= e^{-2i \theta }$ by  (\ref{phasespace}) this formula defines the geodesic trajectory as
\be\label{coor1}
x'=x'(x,y,u,t),~~~y'=y'(x,y,u,t),
\ee
and calculating  the ratio $ u'= {\overline{\omega'_2} / \omega'_2}  = e^{-2i \theta' }$
we shall get 
\be\label{evolu2}
u' = 
{u \cosh( t/2 )  + \sinh( t/2 )   \over  \cosh( t/2 )    + u \sinh( t/2 ) },
\ee 
 that is 
 \be\label{coor2}
u' = u'(u, t). 
 \ee
The equations (\ref{evolu1}) and (\ref{evolu2}) fully describe the time evolution of the phase 
space coordinates $(x,y,u) \in \CM$.

For the motion $g_2(t)$ in (\ref{transformation2}),  $\chi_1 = \cosh( t/2 ) +i \sinh( t/2 )$,   
$\chi_2 = \sinh( t/2 ) +i \cosh( t/2 )$  and we have 
$\omega'_1 =\cosh( t/2 ) \omega_1 + i \sinh( t/2 ) \overline{\omega_1},$~
  $\omega'_2 =\cosh( t/2 ) \omega_2  + i \sinh( t/2 ) \overline{\omega_2}$.
The expression for  $\tau'$ will be the same, except of an additional factor $i$ in front of the $u$.
These expressions allow to define the transformation of the functions  $\{ f(x,y,u)\}$
under the time evolution as $f(x,y,u) \rightarrow  f(x',y',u')$.

The abelian subgroup of $G$ which defines the $SO(2)$ rotations is given by the matrices 
\be
g_{\phi}=\left(
\begin{array}{cc}
\cos \phi & \sin \phi \\
-\sin \phi  & \cos \phi \\
\end{array} \right),~~~~~det g_{\phi} =1.
\ee
For them $\chi_1 = e^{i \phi}$,   $\chi_2 = i e^{i \phi}$  and the transformations they induce are $\omega'_1 =  \omega_1 e^{i\phi},$~  $\omega'_2 =e^{i \phi} \omega_2   $, $\tau' = \tau$.

Using the Stone's theorem this transformation of functions can be 
expressed as an action of a one-parameter 
group of  unitary operators $U_{t}$:
\be\label{geoflow}
U_{t} f(g ) = f(g g_t ).
\ee
Let us calculate  transformations which are induced by $g_1(t)$, $g_2(t)$ and $g_{\phi}$ 
 in (\ref{transformation1})-(\ref{transformation3}). For that purpose 
it is more convenient  to use the {\it independent phase space variables} $(x,y,\theta) \in \CM$ 
or  $(x,y, u)$ and the functions $f(x,y,u)$, where $u = \exp{(-2i \theta)}$. The time evolution of the variables 
 $(x,y,u)$ is given by the equations (\ref{evolu1}) and (\ref{evolu2}):
\beqa\label{evolu}
&U_1(t) f(x,y,u) = f\Big( { u \cosh( t/2 )  + \sinh( t/2 )   \over  \cosh( t/2 )    + u \sinh( t/2 ) }
, ~
{\tau \cosh( t/2 )  + u ~\overline{\tau} ~\sinh( t/2 )  \over \cosh( t/2 )   + u ~\sinh( t/2 ) } \Big),\nn\\
&U_2(t) f(x,y,u) = f\Big({ u \cosh( t/2 )   -i \sinh( t/2 )   \over  \cosh( t/2 )    + i  u \sinh( t/2 ) }, ~
{\tau \cosh( t/2 )  + i u ~\overline{\tau} ~\sinh( t/2 )  \over \cosh( t/2 )   + i u ~\sinh( t/2 ) } \Big),\nn\\
&U_{\phi} f(\omega_2, \tau) = f\Big( e^{i\phi} \omega_2, ~ \tau \Big).
\eeqa
A one-parameter family of unitary operators $U_t$ can be represented as an exponent of the 
self-adjoint  operator $U_t=\exp(i  H  t)$, thus we have 
\begin{equation}
U_{t} f(g )= e^{i  H  t} f(g ) = f(g g_t )
\end{equation}
and differentiating it over the time $t$ at $t=0$ we shall get 
$
   Hf =-i \frac{ \mathrm d}{ \mathrm dt}U_t f \vert_{t=0}. 
$
Let us calculate operator $H$ corresponding to the  $U_1(t)$ and $U_2(t)$.  The time evolution of these variables was given above in (\ref{evolu}). 
Differentiating over time in (\ref{evolu})  we shall get for $H_1$ and $H_2 $:
    \begin{eqnarray}
   2  H_1  =\frac{y }{u}\left(\frac{\partial }{\partial x}+i \frac{\partial }{\partial y}\right)-i \frac{\partial }{\partial u} - u y \left( \frac{\partial }{\partial x}-i\frac{\partial }{\partial y}\right) 
+i u^2 \frac{\partial }{\partial u}  \nn\\
 2 i H_2  =\frac{y }{u}\left(\frac{\partial }{\partial x}+i \frac{\partial }{\partial y}\right)-i \frac{\partial }{\partial u} + u y \left( \frac{\partial }{\partial x}-i\frac{\partial }{\partial y}\right)
-i u^2 \frac{\partial }{\partial u} . 
    \end{eqnarray}
Introducing annihilation and creation operators  $ H_{-}=H_1 -iH_2 $ and  $ H_{+}=H_1 +iH_2 $
 yields 
\begin{eqnarray}
&H_{+}=
\frac{y }{u}\left(\frac{\partial }{\partial x}+i \frac{\partial }{\partial y}\right)-i \frac{\partial }{\partial u},~~~ 
&H_{-}=- u y \left( \frac{\partial }{\partial x}-i\frac{\partial }{\partial y}\right)
+i u^2 \frac{\partial }{\partial u} 
\end{eqnarray}
and calculating the commutator $[H_+, H_-]$ we shall get 
$
H_{0}=u \frac {\partial }{\partial u}$
and  their  $sl(2,R)$ algebra is:
\begin{eqnarray}
\left[H_+,H_-\right]=2 H_0, ~~
\left[H_0,H_+\right]=-H_+,~~
\left[H_0,H_-\right]=H_-.
\end{eqnarray}
We shall consider a class of functions which fulfil the following two equations: 
\begin{eqnarray}
H_0 f(x,y,u) = -{n \over 2}  f(x,y,u) ,  ~~~~~
    H_{-} f(x,y,u)=0 ,  
    \end{eqnarray}
where $n$ is an integer number. The first equation has  the solution $f_n(x,y,u)= ({ 1  \over  u y})^{n/2} \psi(x,y)=\omega^n_2 \psi(x,y)$ and substituting it into the second one we shall get 
\be
n \psi(\tau,\overline{ \tau }) +(\overline{ \tau }-\tau)\frac{\partial \psi(\tau,\overline{ \tau })}{\partial \tau } =0.
\ee
Taking  $\psi(\tau,\overline{ \tau }) = (\overline{ \tau }-\tau)^n \Theta(\overline{ \tau },\tau) $ we shall get 
the equation $\frac{\partial \Theta}{\partial \tau } =0 $, that is $\Theta$
is a anti-holomorphic function and  $f(\omega_2,\tau,\overline{ \tau })$ takes the 
form\footnote{The factors  $(2 i)^n$ have been  absorbed  by the redefinition of $\Theta$.}
 \begin{eqnarray}\label{classfunc}
     f(\omega_2,\tau,\overline{ \tau })=\omega^n_2 (\overline{\tau}-\tau)^n \Theta
     (\overline{\tau}) = {1 \over \overline{\omega_2}^n }\Theta(\overline{\tau}).
     \end{eqnarray}
The invariance under the action of the modular transformation (\ref{periodic}) will take 
the form
\be 
{1\over (p  \overline{\omega_1} + q  \overline{\omega_2} )^n} \Theta ({\frac{m \overline{\tau} + n  }{p \overline{\tau} + q  }})=  {1\over \overline{\omega_2}^n}  \Theta(\overline{\tau}).
\ee
Therefore $\Theta(\overline{\tau})$  is a Poincar\'e theta-function of weight $n$ \cite{Poincare,Poincare1,Ford}:
\be
\Theta ({\frac{m \overline{\tau} + n  }{p \overline{\tau} + q  }}) = \Theta(\overline{\tau}) 
(p \overline{\tau} + q )^n .
\ee
The operator $H_+$ will create the theta function of weight $n+1$.

The invariant integration measure on the group $G$ is given by the formula 
$
 d\mu =dg_{12 } dg_{21} dg_{22} / g_{22}
$
and using (\ref{para}) and (\ref{ratios}) one can express measure in terms of the independent  parameters $(x,y,\theta)$
\begin{equation}
x=-\frac{g_{21}}{g_{22}((g_{21})^2+(g_{22})^2)}
-\frac{g_{12}}{g_{22}},~~~y=\frac{1}{(g_{21})^2 +(g_{22})^2},~~~
\theta =\arctan{ {g_{22} \over  g_{21} } } \nn
\end{equation}
it will take the form \cite{Hopf,Gelfand}
\begin{equation}
d \mu = \frac{dx dy}{y^2} d\theta, 
\end{equation}
which is consistent with the measure defined in our previous section (\ref{me}). 
Thus the invariant integration of functions/observables  
$f(\omega_2,\overline{ \tau },\tau) =  f(\theta, x,y)$
on our  phase space $(x,y,\theta) \in \CM$  will be given by the integral 
\beqa
(f_1 ,f_2)
&=& \int_{0}^{2\pi} d \theta \int_{\CF}f_1(\theta,x,y) \overline{f_2(\theta,x,y) }
{d x d y \over y^2}.
\eeqa
It was demonstrated that the functions on the phase space are of the form (\ref{classfunc})
where $\tau = x + i y$ and ${(\tau-\overline{\tau})\over 2 i }= y$ thus  the expression for 
the scalar product  will takes the following form: 
\beqa
(f_1 ,f_2)
&=& \int_{0}^{2\pi} d \theta \int_{\CF}f_1(\theta,x,y) \overline{f_2(\theta,x,y) }
{d x d y \over y^2} 
=\int_{0}^{2\pi} d \theta \int_{\CF} \Theta_1 (\overline{\tau})  ~  \overline{\Theta_2 (\overline{\tau})}  {1 \over   \vert \omega_2 \vert^{2n} }
  {d x d y \over y^2}=\nn\\
 &=& \int_{0}^{2\pi} d \theta \int_{\CF} \Theta_1 (\overline{\tau})  ~  \overline{\Theta_2 (\overline{\tau})}  (\tau-\overline{\tau})^n   
  {d x d y \over y^2}=  
  \int_{0}^{2\pi} d \theta \int_{\CF} \Theta_1 (\overline{\tau})  ~  \overline{\Theta_2 (\overline{\tau})}  y^n
  {d x d y \over y^2}= \nn\\
  &=&  2\pi  \int_{\CF} \Theta_1 (\overline{\tau})  ~  \overline{\Theta_2 (\overline{\tau})}  y^{n-2}
  d x d y ,~~~~n \geq 2.
\eeqa
This expression for the scalar product allows to calculate the correlations functions. In  the next 
section we shall calculate the correlation functions for  the geodesics (\ref{transformation1})
-(\ref{transformation3}). 

\section{\it Two Point  Correlation Functions  }

The earlier investigation of the correlation functions of  Anosov geodesic flows
was performed in 
\cite{Collet,Pollicot,moore,dolgopyat,chernov} using different approaches including 
Fourier series for the $SL(2, R)$ group, zeta function for the geodesic flows, 
relating the poles of the Fourier transform of the correlation functions to the spectrum of an associated Ruelle operator,  the methods of unitary representation theory,  spectral properties of the 
corresponding Laplacian and others.  In our analyses we shall use the time evolution equations,
the properties of automorphic functions on $\CF$ and shall estimate a decay exponent in terms of 
the phase space curvature and the transformation properties of the functions. 

A  correlation function can be defined as an integral over a pair of  functions/observables 
in which the first one is stationary and the second one evolves with the 
geodesic flow:
\beqa
\CD_{t}(f_1,f_2) &=&   \int_{\CM}f_1(g) \overline{f_2(g g_t) } d \mu 
\eeqa
Using (\ref{coor1}) and (\ref{coor2}) one can represent the integral in a form which is useful 
for our subsequent calculations   \cite{Savvidy:1982jk}:
\beqa
\CD_{t}(f_1,f_2) &=&    
= \int_{0}^{2\pi}\int_{\CF}f_1[x,y,\theta] ~\overline{f_2[x'(x,y,\theta,t), y'(x,y,\theta,t), \theta'(\theta,t)] }
{d x d y \over y^2}d \theta.
\eeqa
In accordance with the previous calculations   (\ref{geoflow}), (\ref{evolu})  and (\ref{evolu1}), (\ref{evolu})
\beqa
&f_1(\omega_2,\tau,\overline{ \tau }) = {1 \over  \overline{\omega_2}^n  } \Theta_1 (\overline{\tau}),~~~\\
&\overline{f_2(\omega'_2,\tau',\overline{ \tau' })} ={1 \over   \omega_2^n  (\cosh( t /2)  +  {\overline{\omega_2}  \over \omega_2}  \sinh( t /2) )^n }  
\overline{ \Theta_2 \Big( {\overline{\tau} \cosh( t/2 )  + {   \omega_2 \over \overline{\omega_2}} ~\tau ~\sinh( t/2 ) 
 \over \cosh( t/2 )   + {   \omega_2 \over \overline{\omega_2}} ~\sinh( t/2 )  }\Big)}.\nn
\eeqa
Therefore the correlation function takes the following form:
\beqa
\CD_t(f_1,f_2)&=&    \int_{0}^{2\pi} d \theta \int_{\CF}  \Phi_1 (\overline{\tau})  \overline{\Phi_2 (\overline{\tau'})} ~ {y^n \over    (\cosh( t/2 )   + {\overline{\omega_2} \over \omega_2} ~\sinh( t/2 )  )^n }  {d x d y \over y^2}=\\
&=&   \int_{0}^{2\pi} d \theta \int_{\CF}  \Theta_1 (\overline{\tau})  \overline{\Theta_2 \Big({ \overline{\tau} e^{ - 2i\theta} \cosh( t/2 )  +  ~\tau ~\sinh( t/2 )   \over
e^{-2i\theta} \cosh( t/2 )   +  ~\sinh( t/2 )  }\Big)} ~ {y^{n-2 } d x d y \over    (\cosh( t/2 )   +  e^{-2i\theta} ~\sinh( t/2 )  )^n }  .  \nn
\eeqa
 In order to calculate the correlation function on arbitrary trajectory we have to
consider the mapping  $g g_t$ given in (\ref{transformation3}). 
In that case for the $\chi_1 , \chi_2 $ we shall get
  \begin{equation}
   \chi_1 =\alpha \exp(t/2) + i \beta \exp(-t/2)  ,~~~  \chi_2 = \gamma \exp(t/2) +  i \delta \exp(-t/2) \nn
  \end{equation}  
and for
\beqa
 \omega'_1 = \omega_1{(\alpha - i \gamma)e^{t/2} + (\delta +i \beta) e^{-t/2} \over 2} +
 \overline{\omega_1}{(\alpha + i \gamma)e^{t/2} - (\delta -i \beta) e^{-t/2} \over 2} ,\nn\\~~~
  \omega'_2 =\omega_2{(\alpha - i \gamma)e^{t/2} + (\delta +i \beta) e^{-t/2} \over 2} +
 \overline{\omega_2}{(\alpha + i \gamma)e^{t/2} - (\delta -i \beta) e^{-t/2} \over 2}, \nn
  \eeqa
thus
\beqa\label{tau}
\tau' ={\omega'_1 \over \omega'_2}=  
  {\tau [(\alpha - i \gamma)e^{t/2} + (\delta +i \beta) e^{-t/2}]  +
u ~\overline{\tau}~  [(\alpha + i \gamma)e^{t/2} - (\delta -i \beta) e^{-t/2} ]   \over
  [(\alpha - i \gamma)e^{t/2} + (\delta +i \beta) e^{-t/2}]  +
u  ~ [(\alpha + i \gamma)e^{t/2} - (\delta -i \beta) e^{-t/2} ] }.
\eeqa
Let us also  consider the  time evolution of  the direction $u'(t)$ of the velocity vector  in a way similar to the  (\ref{evolu1}) and (\ref{evolu2}) , thus
\be
u' = {   [(\alpha + i \gamma)e^{t/2} + (\delta -i \beta) e^{-t/2}] u  +
~ [(\alpha - i \gamma)e^{t/2} - (\delta +i \beta) e^{-t/2} ] 
 \over 
 [(\alpha - i \gamma)e^{t/2} + (\delta +i \beta) e^{-t/2}]  +
 u  ~ [(\alpha + i \gamma)e^{t/2} - (\delta -i \beta) e^{-t/2} ]}
\ee 
and as the time tends to infinity $t \rightarrow \pm \infty$ we shall get 
\be
u'=e^{-2i \theta' }  \rightarrow \pm 1, ~~~~\text{thus} ~~~~~~    \theta' \rightarrow  0 , \pi , 2\pi;
{ \pi \over 2}, {3\pi \over 2}~.
\ee
We can now calculate the correlation function for general geodesic flow:
\beqa
\CD_t(f_1,f_2) &=& \int_{0}^{2\pi} d \theta \int_{\CF}f_1(\omega_2,\tau,\overline{ \tau }) \overline{f_2(\omega'_2,\tau', \overline{ \tau' }) }
{d x d y \over y^2}\\
&=&   \int_{0}^{2\pi} d \theta \int_{\CF}  \Theta_1 (\overline{\tau})  \overline{\Theta_2 (\overline{\tau'})} ~ {y^{n-2}  d x d y\over    [(\alpha - i \gamma)e^{t/2} + (\delta +i \beta) e^{-t/2}]  +
u   ~ [(\alpha + i \gamma)e^{t/2} - (\delta -i \beta) e^{-t/2} ]^{n} }  \nn\\
&=&   \int_{0}^{2\pi} d \theta \int_{\CF}  \Theta_1 (\overline{\tau})  \overline{\Theta_2 (\overline{\tau'})} ~ {y^{n-2}  d x d y\over    [(\alpha - i \gamma)e^{t/2} + (\delta +i \beta) e^{-t/2}]  +
e^{-2i \theta }  ~ [(\alpha + i \gamma)e^{t/2} - (\delta -i \beta) e^{-t/2} ]^{n} }, \nn
\eeqa
where the $\tau'$ is given by (\ref{tau}) and from (\ref{phasespace}) we have $ {\overline{\omega_2} / \omega_2}  
= e^{-2i \theta }$. 
We are interested in finding the upper bound on the correlations functions, thus 
\beqa
&\vert \CD_t(f_1,f_2) \vert   \leq   ~ \\
&   \int_{0}^{2\pi} d \theta \int_{\CF}  \vert    \Theta_1 (\overline{\tau})  \overline{\Theta_2 (\overline{\tau'})}  \vert ~\vert   {y^{n-2}  d x d y\over    [(\alpha - i \gamma)e^{t/2} + (\delta +i \beta) e^{-t/2}]  +
e^{-2i \theta }  ~ [(\alpha + i \gamma)e^{t/2} - (\delta -i \beta) e^{-t/2} ]^{n} } \vert .~ \nn
\eeqa
The absolute value of the denominator is 
$
4 e^{t}( \alpha \cos \theta + \gamma \sin \theta)^2 + 4 e^{-t}( \beta \cos \theta + \delta \sin \theta)^2,\nn
$
therefore in the limit $t \rightarrow + \infty$
\beqa
\vert \CD_t(f_1,f_2) \vert  & \leq &   ~    \int_{0}^{2\pi} d \theta \int_{\CF}  \vert   \Theta_1 (\overline{\tau})  \overline{\Theta_2 (\overline{\tau'})}  \vert ~  {y^{n-2}  d x d y \over   2^n [ e^{t}( \alpha \cos \theta + \gamma \sin \theta)^2 +  e^{-t}( \beta \cos \theta + \delta \sin \theta)^2 ]^{{n\over 2}} } =~ \nn\\
& = &\int_{0}^{2\pi} d \theta \int_{\CF}  \vert    \Theta_1 (\overline{\tau})  \overline{\Theta_2 (\overline{\tau'})}  \vert ~   {y^{n-2}  d x d y \over   2^n [ ( \alpha \cos \theta + \gamma \sin \theta)^2 +  e^{-2t}( \beta \cos \theta + \delta \sin \theta)^2 ]^{{n\over 2}} } ~ e^{-{n\over 2} t} .~\nn\\~~
\eeqa
Discarding the positive term $e^{-2t}( \beta \cos \theta + \delta \sin \theta)^2$ in the denominator 
we shall get 
\beqa
\vert \CD_t(f_1,f_2) \vert  & \leq &   
 \int_{0}^{2\pi} d \theta \int_{\CF}  \vert   \Theta_1 (\overline{\tau})  \overline{\Theta_2 (\overline{\tau'})}  \vert ~   {y^{n-2}  d x d y \over   2^n [( \alpha \cos \theta + \gamma \sin \theta)^{2}]^{{n\over 2}}} ~ e^{-{n\over 2} t}.~\nn
\eeqa
The geodesic flow is given by the evolution equation (\ref{transformation3}), therefore taking the limits 
$t \rightarrow \pm \infty$ one can express the parameters of the matrix $g=g(\alpha,\beta,\gamma,\delta)$ 
in terms of the coordinates $(\xi,\eta)$  in (\ref{region}):
\be
z(-\infty) = \eta = {\beta \over \delta},~~~~z(+\infty) = \xi = {\alpha \over \gamma}  > 1,\nn
\ee
thus 
\beqa
\vert \CD_t(f_1,f_2) \vert  & \leq &   {1\over  2^n \vert \gamma \vert^n}
 \int_{0}^{2\pi} d \theta \int_{\CF}  \vert    \Theta_1 (\overline{\tau})  \overline{\Theta_2 (\overline{\tau'})}  \vert ~   {y^{n-2}  d x d y \over   [( \xi \cos \theta +  \sin \theta)^{2}]^{{n\over 2}}} ~ e^{-{n\over 2} t}.~
\eeqa
The last integral is an average over all geodesic trajectories which are defined by the position  $(x,y) \in \CF $ and the directions of the velocity vectors $\vec{v} =(\cos \theta, \sin \theta)$.   In order to exclude the apparent singularity at the angle $\theta_0$ which solves the equation 
$\xi \cos \theta_0 + \sin \theta_0 =0 $ we shall exclude small $\epsilon$ volume of the phase space 
which is surrounding the singular direction $\theta_0$.  With such an assumption we shall get 
an exponential bound on the correlation functions of the form 
\beqa
\vert \CD_t(f_1,f_2) \vert  & \leq &     ~ M_{\Theta_1 \Theta_2}(\epsilon)~  e^{-{n\over 2} t}~~.
\eeqa
In the opposite limit $t \rightarrow - \infty$  we shall get 
\beqa
\vert \CD_t(f_1,f_2) \vert  & \leq &   {1\over  2^n \vert \delta \vert^n}
 \int_{0}^{2\pi} d \theta \int_{\CF}  \vert    \Theta_1 (\overline{\tau})  \overline{\Theta_2 (\overline{\tau'})}  \vert ~   {y^{n-2}  d x d y \over   [( \eta \cos \theta +  \sin \theta)^{2}]^{{n\over 2}}} ~ e^{{n\over 2} t}.~
\eeqa
We can combine the results obtained in the both limits  $t \rightarrow \pm \infty$  into the
following bound 
\beqa
\vert \CD_t(f_1,f_2) \vert  & \leq &     ~ M_{\Theta_1 \Theta_2}(\epsilon)~  e^{-{n\over 2}\vert  t \vert }~~.
\eeqa
If the surface has a negative curvature $K $ 
\be
ds^2 = {dx^2 +dy^2 \over K  y^2},
\ee 
then in the last formula the exponential factor  will take the form 
\beqa
\vert \CD_t(f_1,f_2) \vert  & \leq &     ~ M_{\Theta_1 \Theta_2}(\epsilon)~  e^{-{n\over 2} K \vert t \vert}~~.
\eeqa
Introducing the characteristic time decay   \cite{yer1986a,Savvidy:2018ygo} we shall get
\be
\tau_0 = {2\over n K }.
\ee
The decay time  of the correlation functions 
is shorter when the surface has larger negative curvature, or in other words when the divergency  
of the trajectories is stronger.

\section{\it Acknowledgment }

This project has received funding from the European Union's Horizon 2020 research and innovation programme under the Marie Sk\'lodowska-Curie grant agreement No 644121.
G.S. would like to thank Spenta Wadia for pointing out  the 
article of Emil Artin  \cite{Artin}.

\section{\it Appendix A}
The rate of convergence at which the continued fractions approximate the real numbers can be 
understood  in terms of the ratios $\frac{P_{n}}{Q_{n}}$.  
From (\ref{PQrel}) we get
\be\label{P_nQ_nfrac}
\frac{P_{n}}{Q_{n}}=
\frac{P_{n-1}a_{n-1}+P_{n-2}}{Q_{n-1}a_{n-1}+Q_{n-2}}. 
\ee
Let us show first that one can obtain  the ratio $\frac{P_{n+1}}{Q_{n+1}}$ from $\frac{P_{n}}{Q_{n}}$
by making the replacement 
\be \label{repl_an}
a_{n-1}\to a_{n-1} +\frac{1}{a_{n}}\,.
\ee
Indeed, substituting the replacement (\ref{repl_an})  into the (\ref{P_nQ_nfrac}) we shall get
\beqa
&\frac{P_{n}}{Q_{n}}=
\frac{P_{n-1}a_{n-1}+P_{n-2}}{Q_{n-1}a_{n-1}+Q_{n-2}} \to \\ \nonumber
&\frac{P_{n-1}(a_{n-1} +\frac{1}{a_{n}})+P_{n-2}}{Q_{n-1}(a_{n-1} +\frac{1}{a_{n}})+Q_{n-2}}=
\frac{(P_{n-1}a_{n-1}+P_{n-2}) a_n+P_{n-2}}{(Q_{n-1}a_{n-1}+Q_{n-2}) a_n+Q_{n-2}}=
\frac{P_{n+1}}{Q_{n+1}},
\eeqa
where the last step is a consequence of (\ref{PQrel}). 

On the other hand, from 
(\ref{P_nQ_nfrac}) and (\ref{P_0,Q_0}) it follows that $\frac{P_1}{Q_1}=a_0$ and we can construct 
the ratios $\frac{P_2}{Q_2},\frac{P_3}{Q_3},\dots ,\frac{P_n}{Q_n}$ using the recurrence 
(\ref{repl_an}). The result will take the following form: 
\begin{small}
\beqa
\frac{P_1}{Q_1}=a_0, \,\,
\frac{P_2}{Q_2}=a_0+\frac{1}{a_1},\,
\frac{P_3}{Q_3} = a_0 + \cfrac{1}{a_{1} +\cfrac{1}{a_{2}} }\,,
\dots\,, 
\frac{P_n}{Q_n}=
a_0+\cfrac{1}{a_1 +\cfrac{1}{a_2 +\cfrac{1}{a_3 + \lastcfrac{1}{a_{n-1}}}}}\,.
\eeqa
\end{small}
 Because of the condition (\ref{PQrecrel})
$$
P_n Q_{n-1} - Q_n P_{n-1} =(-1)^n~
$$
it follows that when $n$ is even the ratios $\frac{P_n}{Q_n} $ are increasing, but  are  always less than $\xi$ and 
for odd $n$ the ratios $\frac{P_n}{Q_n} $ are decreasing and are  always greater than $\xi$, therefore 
\be
\left|\xi -\frac{P_n}{Q_n}\right|<
\left|\frac{P_{n+1}}{Q_{n+1}} -\frac{P_n}{Q_n}\right|<
\frac{1}{Q_{n}Q{n+1}}<\frac{1}{Q_n^2}. 
\ee
The sequence of denominators $Q_n$ is increasing function of $n$, as it follows from (\ref{PQrel}), (\ref{P_0,Q_0}) and from the fact that the $a_n$s are nonzero positive integers: 
$Q_{n+1}>Q_{n}$. Thus every real number $\xi$  can be approximated by  the ratio $\frac{P_n}{Q_n}$ as:
\be
\left|\xi -\frac{P_n}{Q_n}\right|<\frac{1}{Q_n^2}. 
\ee
and obviously 
$$
\lim_{n\to\infty}\frac{P_n}{Q_n} \to \xi~.
$$
The ratio $\frac{P_n}{Q_n}$ is called the $n$th convergent. 

 \section{\it Appendix B}
 
One can use alternative independent variables $(u, \tau, \bar{\tau})$ instead of $(u,x,y)$ in order to describe a geodesic flow. In that case the operators $(H_+,H_-,H_0)$   will take an alternative form.  In some cases 
this representation  has an advantage to be more appropriate for operator calculations. We shall present them below. The transformation of functions will take the form:
\beqa
&U_1 f(u,\tau,\bar{\tau})=
f\left(\frac{u \cosh \left(\frac{t}{2}\right)+\sinh \left(\frac{t}{2}\right)}{u \sinh \left(\frac{t}{2}\right)+\cosh \left(\frac{t}{2}\right)},\frac{\bar{\tau} u \sinh \left(\frac{t}{2}\right)+\tau \cosh \left(\frac{t}{2}\right)}{u \sinh \left(\frac{t}{2}\right)+\cosh \left(\frac{t}{2}\right)},\frac{\bar{\tau} \cosh \left(\frac{t}{2}\right)+\frac{\tau}{u}\sinh \left(\frac{t}{2}\right)}{\frac{1}{u}\sinh \left(\frac{t}{2}\right)+\cosh \left(\frac{t}{2}\right)}\right) ,\nonumber \\
&U_2 f(u,\tau,\bar{\tau})=
f\left(\frac{u \cosh \left(\frac{t}{2}\right)-i \sinh \left(\frac{t}{2}\right)}{\cosh \left(\frac{t}{2}\right)+i u \sinh \left(\frac{t}{2}\right)},\frac{\tau \cosh \left(\frac{t}{2}\right)+i \bar{\tau} u \sinh \left(\frac{t}{2}\right)}{\cosh \left(\frac{t}{2}\right)+i u \sinh \left(\frac{t}{2}\right)},\frac{\bar{\tau} \cosh \left(\frac{t}{2}\right)-i \frac{ \tau}{u}\sinh \left(\frac{t}{2}\right)}{\cosh \left(\frac{t}{2}\right)-i\frac{1}{u} \sinh \left(\frac{t}{2}\right)}\right).\nn
\eeqa
By applying 
\begin{equation}
Hf =-i \frac{ \mathrm d}{ \mathrm dt}U_t f \vert_{t=0}\nn
\end{equation}
we will get
\beqa
&H_1=-i\frac{1}{2}  \left(-\left(u^2-1\right) \frac{\partial}{\partial u}+\frac{(\tau-\bar{\tau}) }{u}\frac{\partial}{\partial \bar{\tau}}+u (\bar{\tau}-\tau) \frac{\partial}{\partial \tau}\right),\\
&H_2=\frac{1}{2}   \left(\frac{(\bar{\tau}-\tau) }{u}\left(u^2 \frac{\partial}{\partial \tau}+\frac{\partial}{\partial \bar{\tau}}\right)-\left(u^2+1\right) \frac{\partial}{\partial u}\right).
\eeqa
After defining the $ H_{-}=H_1 -iH_2 $ and  $ H_{+}=H_1 +iH_2 $ we shall get:
\beqa
&H_{+}=-i \left(\frac{(\tau-\bar{\tau})}{u} \frac{\partial }{\partial \bar{\tau}}+\frac{\partial }{\partial u}\right)\,,\\
&H_{-}=i u \left((\tau-\bar{\tau}) \frac{\partial }{\partial \tau}+u \frac{\partial }{\partial u}\right)\,,\\
&H_0=u \frac{\partial}{\partial u}
\eeqa
and the corresponding algebra is:
\beqa
&\left[H_0,H_{+}\right]=-H_{+}\,,\nn\\
&\left[H_0,H_{-}\right]=H_{-}\,,\nn\\
&\left[H_{+}H_{-}\right]=2H_0\,. \nn
\eeqa

\end{document}